# Most, And Least, Compact Spanning Trees of a Graph

**Gyan Ranjan**[*]
Ericsson Inc. & Netlabs.ai
California, USA
gyan.ranjan@netlabs.ai

**Nishant Saurabh**
Dept. of Information & Computing Sciences
Utrecht University, The Netherlands
n.saurabh@uu.nl

**Amit Ashutosh**[*]
Cisco Systems Inc.
California, USA
amashuto@cisco.com

June 17, 2022

## Abstract

We introduce the concept of **Most, and Least, Compact Spanning Trees** - denoted respectively by $T^*(G)$ and $T^\#(G)$ - of a *simple, connected, undirected and unweighted* graph $G(V, E, W)$. For a spanning tree $T(G) \in \mathcal{T}(G)$ to be considered $T^*(G)$, where $\mathcal{T}(G)$ represents the set of all the spanning trees of the graph $G$, it must have *the least average inter-vertex pair (shortest path) distances* from amongst the members of the set $\mathcal{T}(G)$. Similarly, for it to be considered $T^\#(G)$, it must have the *highest average inter-vertex pair (shortest path) distances*. In this work, we present an iteratively greedy *rank-and-regress* method that produces at least one $T^*(G)$ or $T^\#(G)$ by eliminating one extremal edge per iteration. The rank function for performing the elimination is based on the elements of the matrix of *relative forest accessibilities* of a graph and the related *forest distance* [22]. We provide empirical evidence in support of our methodology using some standard graph families: complete graphs, the Erdős-Renyi random graphs and the Barabási-Albert scale-free graphs; and discuss computational complexity of the underlying methods which incur polynomial time costs.

## 1 Introduction

Compact sub-structures (sub-graphs) of graphs find use in many scientific disciplines: from resource placement [43] to routing [30, 31], structural analysis to data compression [45, 46], information coding and machine learning [15, 19], to name but a few. One of the most popular and widely studied compact sub-structures of a graph are the spanning trees; and, in particular, for weighted graphs - and digraphs (directed graphs) - the *Minimum Spanning Trees* (MST) [48]. For a spanning tree to be considered an MST, it must have the least sum of edge weights from amongst all the spanning trees of the graph. Several well known algorithms exist for extracting MSTs of a graph; the Prim's and Kruskal's algorithms being the two most widely used [48]. Both algorithms follow a *greedy* selection approach towards constructing the MST; induced by a rank order defined over the edge weights. Alas, when a graph $G$ is unweighted - i.e. all edges have a unit weight - every spanning tree of $G$ is an MST, with a cumulative weight of $n-1$, $n$ being the number of nodes/vertices in the graph. This renders the utility of MST as a compact sub-structure rather meaningless.

Another class of compact spanning sub-graphs is the so-called *shortest path tree* [1, 3, 5], which seems to have been used a lot in the field of communication networks and distributed systems [2, 4]. A shortest path tree is a spanning sub-graph which is constructed by choosing a vertex as root, and ensuring the least possible shortest path distance between it and the rest of the vertices in the graph. The optimization of path lengths is with respect to the root vertex and the trees are not necessarily the most compact spanning acyclic sub-graphs of the graph.

---

[*]The authors are currently employed at Ericsson, USA; and Cisco Systems, USA, respectively. However, this research has been conducted entirely in personal capacity based on individual interests. Ericsson and Cisco, neither bear responsibility, nor claim any credit for it.

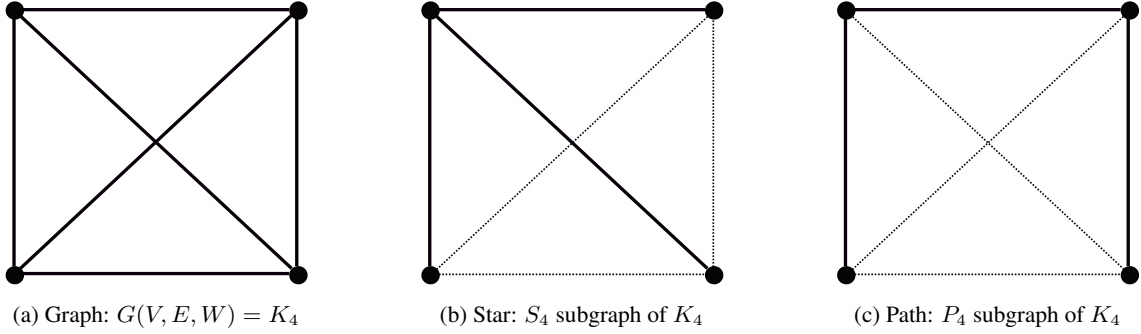

(a) Graph: $G(V, E, W) = K_4$     (b) Star: $S_4$ subgraph of $K_4$     (c) Path: $P_4$ subgraph of $K_4$

Figure 1: (a) The complete graph $K_4$ of order $n = 4$, (b) an $S_4$: one of four MCSTs of $K_4$, (c) a $P_4$: one of twelve LCSTs of $K_4$. Solid lines in (b) and (c) represent the edges of $S_4$ and $P_4$ respectively, while the dotted lines represent edges from $K_4$ that are not retained in the respective trees.

In this work, we introduce and study a new class of spanning acyclic sub-graphs: the *Most* and the *Least Compact Spanning Trees*. For brevity, denoted henceforth as MCST ($T^*(G)$) and LCST ($T^\#(G)$), respectively. For a spanning tree $T(G) \in \mathcal{T}(G)$ to be considered $T^*(G)$ - where $\mathcal{T}(G)$ represents the set of all the spanning trees of the graph $G$ - it must have the *least average inter-vertex pair (shortest path) distances* from amongst the members of the set $\mathcal{T}(G)$. On the other hand, for it to be considered $T^\#(G)$, it must have the *highest average inter-vertex pair (shortest path) distances*. Clearly, neither $T^*(G)$ nor $T^\#(G)$ are necessarily unique i.e. there can be more than one of each in a given graph $G$. For example, consider a graph $G$ in which at least one vertex, say $v_i \in V(G)$, has a degree $d(v_i) = |V(G)| - 1 = (n - 1)$. Clearly, the star graph $S_n$ rooted at $v_i$ represents one of the MCST of $G$. More than one such vertices/nodes may exist in $G$. Similarly, if the path graph ($P_n$) is a sub-graph of $G$, then it is an LCST of $G$. More than one paths of length $(n - 1)$ may exist in $G$ (e.g. in a graph with a fundamental cycle $C_n$ of order $n$).

The MCST and LCST algorithms use the inter-vertex pair *Forest Distance* [22], limited to node pairs connected by an edge, and eliminate one edge per iteration until a spanning tree results in the end i.e. $(n - 1)$ edges remain. No *bridge* edge - an edge that when deleted renders the remaining graph disconnected - ever gets deleted during the process. This is done by using the *effective resistance distance* which - when measured across the end-points of a bridge edge, is always 1 (for an unweighted graph). The *iteratively greedy* nature of our algorithm is a result of the fact that the rank order of edges needs to be recomputed in each iteration, unlike the universal edge weight based rank order used by the Prim's and Kruskal's algorithms for extracting MSTs [48].

The rest of this work is organized into the following sections: we start by formally stating the problem, along with all the notations and paraphernalia in §2. In §3, we cover the *matrix of relative forest accessibilities* of a graph, and the associated *forest distance*, followed by our *iteratively greedy rank-and-regress* algorithm based on the elements corresponding to the set of edges in the graph. Next, in §4, we present thorough empirical evaluations using several standard graph families, and establish that the iteratively greedy rank-and-regress method indeed produces the MCST ($T^*(G)$) and LSCT ($T^\#(G)$) respectively, for the two extremal choices. This is followed up, in §5, with a discussion on the computational complexity and various ways in which the costs can be optimized. We review related work literature in §6. And finally, in §7, we conclude with a discussion of future work and potential research directions.

## 2 Problem Formulation

In this section, we formally pose the problem of extracting an MCST ($T^*(G)$) and an LCST ($T^\#(G)$). But first a few notations and preliminaries are in order.

### 2.1 Notations and Preliminaries

#### 2.1.1 Topological Primitives

We denote by $G(V, E, W)$ a **simple, connected, undirected and unweighted graph**; where the set of vertices/nodes of $G$ are represented by $V(G)$ and the set of edges/links by $E(G)$. The number of vertices $|V(G)| = n$ is also called the *order* of the graph $G$ and the number of edges $|E(G)| = m$ is related to the *volume* of the graph (defined in the next paragraph) [2].

---

[2] In the rest of this work we use the terms vertices and nodes interchangeably; the same goes for edges and links.

We denote by $v_i \in V(G)$ the $i^{th}$ vertex of $G$ (the ordinality is arbitrary); and an edge connecting $v_i$ and $v_j$ by $e_{ij}$. As the graph is unweighted, all edges have a unit weight, i.e. $w_{ij} = 1$ denotes the weight of edge $e_{ij}$. The quantity $d(v_i)$ denotes the degree of the vertex $v_i$ and represents the count of vertices that $v_i$ is directly connected to through an edge. Note that $Vol(G) = \sum_{i=1}^{n} d(v_i) = 2m$; is the aforementioned *volume*. Finally, $\mathcal{T}(G)$ represents the set of all the spanning trees of $G$.

### 2.1.2 Algebraic Paraphernalia

Next, we introduce all the algebraic functions that help us design the metrics required in our algorithms. By $A \in \Re^{n \times n}$, we denote the adjacency matrix of $G(V, E, W)$ [12], such that:

$$A_{ij} = 1 \quad if \ e_{ij} \in E(G),$$
$$= 0 \quad Otherwise.$$

Clearly, $A$ is a square symmetric matrix. The *combinatorial Laplacian* of the graph $G(V, E, W)$, is defined as:

$$L = D - A \tag{1}$$

where $D = [d_{ii}] = d(v_i)$, is a diagonal matrix with the node degrees on the diagonal; and 0 elsewhere.

$L$ is a square, symmetric, doubly-centered (all rows and columns sum up to 0) and positive semi-definite matrix with $n$ non-negative eigen values [36, 37, 38]. It's eigen vectors form the orthonormal basis of order $n$. $L$ is not invertible, owing to a unique minimal eigen value 0, if the graph $G$ is simple, undirected and connected. The *Moore-Penrose Pseudo-Inverse* of $L$, denoted by $L^+$ is a generalized inverse [11, 17, 32, 34, 35]. Both $L$, in a modified form akin to the Tikhonov Regularization [28], as well as it's pseudo-inverse $L^+$ yield distance functions which we exploit in §3 to construct our *iteratively greedy rank-and-regress* algorithm for generating the most and least compact spanning trees of a given $G$.

## 2.2 Compactness of a Spanning Tree

**Definition 1** *Spanning Tree of a graph: A spanning tree $T(G)$ of the graph $G(V, E, W)$ is a connected acyclic spanning sub-graph of $G$ with exactly $n - 1$ edges in it.*

The term *spanning* applied to a sub-graph simply means that it has all the vertices of the graph $G$ in it. In Figure 1, both $S_4$ and $P_4$ represent spanning trees of $K_4$. As we shall see later, $T(G)$ is also a special kind of spanning forest of $G$ (one with **exactly one** tree in it). A *Minimum Spanning Tree* (MST) of a graph is a spanning tree with the least sum of edge weights across all possible spanning trees. But when $G$ is unweighted, all edges have the same unit weight. Clearly, the volume of any $T(G)$ is given by:

$$Vol(T(G)) = 2(n-1), \qquad \forall T(G) \in \mathcal{T}(G) \tag{2}$$

which is an invariant across the set of spanning trees of a given graph. Therefore, algorithms like Prim's or Kruskal's cannot differentiate between them. In contrast, we define the following as a *compactness* metric that can:

**Definition 2** *The Compactness Metric: Average inter-vertex (shortest path) distances in a graph $G$.*

$$\mathcal{C}(G) = \frac{1}{n^2} \sum_{i=1}^{n} \sum_{j=1}^{n} \mathcal{D}(v_i, v_j) \tag{3}$$

where $\mathcal{D}(v_i, v_j)$ is the length of the path connecting $v_i$ to $v_j$ (or, *vice versa*) in a graph $G$. Given that $G(V, E, W)$ is unweighted, $\mathcal{D}(v_i, v_j)$ is simply the *hop count* from $v_i$ to $v_j$. With this in view, we define the MCST ($T^*(G)$) and LCST ($T^\#(G)$) respectively as:

**Definition 3** *Most Compact Spanning Tree of a graph (MCST):*

$$T^*(G) = \underset{T(G) \in \mathcal{T}(G)}{\operatorname{argmin}} \mathcal{C}(T(G)) \tag{4}$$

**Definition 4** *Least Compact Spanning Tree of a graph (LCST):*

$$T^\#(G) = \underset{T(G) \in \mathcal{T}(G)}{\operatorname{argmax}} \mathcal{C}(T(G)) \tag{5}$$

Note again in Figure 1, the star graph $S_4$ and the path graph $P_4$ respectively constitute an MCST and an LCST in $K_4$. In what follows, we describe an iteratively greedy *rank-and-regress* algorithm that eliminates $[m - (n - 1)]$ edges of a graph to produce at least one $T^*(G)$ (or, $T^\#(G)$; depending upon the nature of the greedy choice (c.f. §3.2 and §4).

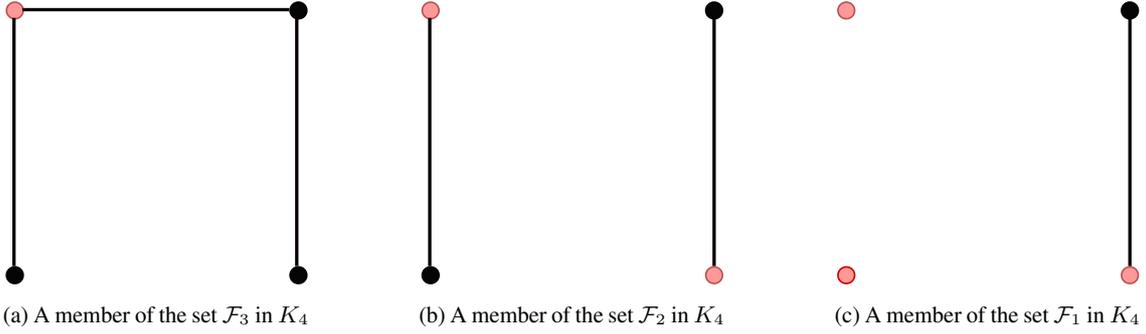

(a) A member of the set $\mathcal{F}_3$ in $K_4$    (b) A member of the set $\mathcal{F}_2$ in $K_4$    (c) A member of the set $\mathcal{F}_1$ in $K_4$

Figure 2: Some spanning rooted forests of $K_4$: (a) a spanning rooted forest with 3 edges (or, one tree), which is the same as a spanning tree, with one root colored in red (top-left), (b) a spanning rooted forest with 2 edges (or, two trees), with one vertex in each tree colored in red (top vertex in the left tree, and bottom vertex in the right), (c) a spanning rooted forest with 1 edge (or, three trees), in which two isolated vertices represent two trees (top left, bottom left), each colored in red as they are roots by default; and one tree on the right, represented by the sole edge, with the bottom right vertex colored in red as root. A member of $\mathcal{F}_0$ in $K_4$ is simply a collection of four isolated vertices (i.e. no edges), and each vertex represents a trivial tree with itself as root. So, each tree of a spanning rooted forest has one node designated as root in it.

## 3 Methodology

### 3.1 Spanning Rooted Forests and a Pair of Distances: $[\Delta_{ij}, \Omega_{ij}]$

In this section, we introduce a pair of distance functions called the *forest distance* and the *effective resistance distance*, based on which we define our *iteratively greedy rank-and-regress* algorithm. But first, a word on a special class of spanning sub-graphs called the *spanning rooted forests* of $G$ is due.

**Definition 5** *Spanning Rooted Forest with $k$ edges*: *A spanning acyclic sub-graph of $G$ with $k$ edges, or, equivalently, $n - k$ trees in it, in which each tree has a designated root vertex.*

Computationally, the number of *spanning rooted forests* of $G$ are enumerated using the elements of a single matrix $Q \in \Re^{n \times n}$:

$$Q = (I + L)^{-1} \quad (6)$$

where $I \in \Re^{n \times n}$ is the *identity* matrix whose columns form the standard orthonormal basis in an n-dimensional space, and $L$ is the *combinatorial Laplacian* of $G$ that was introduced in §2.1.2 [3]. The matrix $Q$ is referred to as the matrix of *relative forest accessibilities* (or, the *forest matrix* for brevity) and has some very interesting properties and applications [20, 21, 22]. For a simple, connected, undirected and unweighted graph $G(V, E, W)$, $Q$ is square, symmetric, positive-definite and doubly stochastic; i.e. every row or column of $Q$ sums up to 1.

Next, we connect the elements of $Q$ to the *spanning rooted forests* in $G$. Let $\mathcal{F}_k^{ij}$ denote the set of spanning rooted forests in $G$ with $k$ edges - or, $n - k$ trees - in which $v_i$ and $v_j$ belong to the same tree and $v_i$ is the *root* of that tree. $\mathcal{F}_k$, similarly, represents the set of all spanning rooted forests in $G$ with exactly $k$ edges (or, $n - k$ trees, each with one root in it). Then,

$$Q_{ij} = \sum_{k=0}^{n-1} \varepsilon(\mathcal{F}_k^{ij}) / \sum_{k=0}^{n-1} \varepsilon(\mathcal{F}_k) = Q_{ji} \quad (7)$$

The operator $\varepsilon(\cdot)$ above simply represents the cardinality (count of the elements) in its operand set, because all edges have a unit weight. For weighted graphs, it would represent a sum of products of edge weights in a forest. For details, the reader is referred to [21, 22]. Of particular interest to us is the *Forest Distance* function, defined over all pairs of vertices $(i, j) \in V(G) \times V(G)$:

$$\Delta_{ij} = Q_{ii} - Q_{ij} - Q_{ji} + Q_{jj} = \Delta_{ji} \quad (8)$$

Finally, we come to the matrix $L^+$, the *Moore-Penrose Pseudo-Inverse* of $L$, that was also introduced in §2.1.2. Analogous to $Q$, is related to the *dense spanning rooted forests* of $G$ [21, 22]. A dense spanning rooted forest contains exactly $(n-1)$ or $(n-2)$ edges in it; or, equivalently, has either one or two trees, respectively. To be specific,

$$L_{ij}^+ = \frac{\varepsilon(\mathcal{F}_{n-2}^{ij}) - \frac{1}{n}\varepsilon(\mathcal{F}_{n-2})}{\varepsilon(\mathcal{F}_{n-1})} = L_{ji}^+ \qquad (9)$$

---

[3] Note the similarity to the Tikhonov regularization [28] here. It is easy to see that the form (I + L) simply shifts all the eigen values of $L$ by 1 and makes it invertible.

Once again, the elements of $L^+$ yield the *effective resistance distance* between any pair of vertices $(i,j) \in V(G) \times V(G)$ [33, 41, 42, 50]:

$$\Omega_{ij} = L_{ii}^+ - L_{ij}^+ - L_{ji}^+ + L_{jj}^+ = \Omega_{ji} \qquad (10)$$

In particular, we use $\Omega_{ij}$ to detect a *bridge edge* - i.e. an edge which upon deletion renders a graph disconnected. If $\Omega_{ij} = 1$ and $e_{ij} \in E(G)$, then it is easy to show that $e_{ij}$ must be a bridge edge [42]. In the next sub-section, we use the values of $\Delta_{ij}$ and $\Omega_{ij}$ to construct our *iteratively greedy rank-and-regress* algorithm for extracting an MCST/LCST.

### 3.2 An Iteratively Greedy Rank-And-Regress Algorithm

| # | Pseudo-Code | Comment |
|---|---|---|
| 1 | *init*: $G(V, E, W)$:<br>$\quad n \leftarrow |V(G)|$;<br>$\quad E(G) \leftarrow G(V, E, W)$; | $G$ is simple, connected, undirected, and unweighted.<br>$G$ is a graph of order $n$.<br>Initialize $E(G)$, the set of edges of $G$. |
| 2 | *while* $|E(G)| > (n-1)$: | Ensure that $G$ is not a tree already. If not, start iteration. |
| 3 | $\quad Q = (I + L)^{-1}; \quad L^+ = pinv(L);$ | Compute the Forest matrix and $L^+$ for current $G$. |
| 4 | $\quad$ *init*: $\Delta \leftarrow [0]; \; \Omega \leftarrow [0];$<br>$\quad\quad \delta^* \leftarrow 0; \; e^* \leftarrow \phi;$<br><br>$\quad$ *for each* $e_{ij}$ in $E(G)$:<br><br>$\quad\quad \Delta_{ij} = Q_{ii} - Q_{ij} - Q_{ji} + Q_{jj}$<br>$\quad\quad \Omega_{ij} = L_{ii}^+ - L_{ij}^+ - L_{ji}^+ + L_{jj}^+$<br><br>$\quad\quad$ *if* $\Omega_{ij} == 1$: *continue*;<br><br>$\quad\quad$ *else if* $\Delta_{ij} > \delta^*$:<br>$\quad\quad\quad \delta^* \leftarrow \Delta_{ij}; \; e^* \leftarrow e_{ij};$<br><br>$\quad$ **end for each** | Initialize the Forest and Resistance distance matrices.<br><br>Iterate over the edges in $G$.<br><br>Compute the Forest distance.<br>Compute the Resistance distance.<br><br>Check if $e_{ij}$ is a *bridge edge*.<br><br>Check if $e_{ij}$ improves on the extremality criteron?<br>$e_{ij}$ is the new candidate for being $e^*$.<br><br>Loop back for the next edge. |
| 5 | $\quad E(G) \leftarrow E(G) - e^*$<br>**end while** | Delete $e^*$ from the graph $G$.<br><br>Loop back to step 2 for the next iteration. |
| 6 | *return* $[T^*(G) \leftarrow E(G)];$ | The remaining $(n-1)$ edges in $E(G)$ are the MCST. |

Table 1: **An iteratively greedy rank-and-regress algorithm for constructing an MCST**: Returns one of the *Most Compact Spanning Trees* of the graph $G$. The variant for finding at least one **LCST** simply requires a reversal in the *extremality* criteria. Instead of $\delta^*$ selecting an edge with maximum $\Delta_{ij}$ for deletion in each iteration (c.f. step 4 above), it will select an edge with the minimum $\Delta_{ij}$.

Table 1 presents the pseudo-code for our *iteratively greedy rank-and-regress* algorithm. Although the code targets the extraction of an MCST ($T^*(G)$), with one slight modification, it would extract an LCST ($T^\#(G)$) instead. Note:

a. **Regress, Not Selection**: Unlike the *Minimum Spanning Tree* algorithms, or, the random/shortest-path spanning tree generation, where the tree is constructed by greedily selecting edges, our method eliminates one *extremal* edge at a time. Hence, the qualifier *regress*.

b. **Number of Iterations**: The algorithm begins by initializing the set $E(G)$ of edges and deleting **exactly one** edge from the graph in each iteration until the number of edges in $E(G)$ reduces to $(n-1)$. Hence, if the number of edges in $G$ originally (at the beginning) is $m$, the algorithm concludes in exactly $m-(n-1)$ iterations.

c. **Maintaining Connectedness**: The graph $G$, through this regress, always stays connected. This is ensured in step 4 by skipping any extremal edge $e_{ij}$ which is also a *bridge* edge at that stage; i.e. $\Omega_{ij}=1$.

d. **The Tree Property**: By ensuring connectedness during all iterations, as stated in [c.] above, and noting the fact that there are exactly $(n-1)$ edges remaining in the end, our algorithm ensures that the resulting $E(G)$ in step 6 is indeed a spanning tree.

e. **The Iteratively Greedy Nature**: The *rank* order of the edges remaining in the graph in each iteration may differ across iterations. This can be observed in steps 3 and 4 where $\{Q, L^+, \Delta, \Omega\}$ are recomputed in each iteration. Hence, the qualifier *iteratively greedy*.

### 3.3 Analysis: Why would it work?

In order to make sense of why the greedy choice on $\Delta_{ij} : e_{ij} \in E(G)$ works for attaining the objectives for MCST and LCST - as defined in (4) and (5) respectively - we first need to understand the constituents of $\Delta_{ij}$. Recall, from (8):

$$\Delta_{ij} = Q_{ii} - Q_{ij} - Q_{ji} + Q_{jj} = \Delta_{ji} \quad (11)$$

The term $[Q_{ii} - Q_{ij}]$, based on the definition in (7), yields:

$$Q_{ii} - Q_{ij} = \sum_{k=0}^{n-1}[\varepsilon(\mathcal{F}_k^{ii}) - \varepsilon(\mathcal{F}_k^{ij})] / \sum_{k=0}^{n-1} \varepsilon(\mathcal{F}_k)$$

$$= \frac{[\varepsilon(\mathcal{F}_0^{ii}) - \varepsilon(\mathcal{F}_0^{ij})] + [\varepsilon(\mathcal{F}_1^{ii}) - \varepsilon(\mathcal{F}_1^{ij})] + ... + [\varepsilon(\mathcal{F}_{n-2}^{ii}) - \varepsilon(\mathcal{F}_{n-2}^{ij})] + [\varepsilon(\mathcal{F}_{n-1}^{ii}) - \varepsilon(\mathcal{F}_{n-1}^{ij})]}{\sum_{k=0}^{n-1} \varepsilon(\mathcal{F}_k)}$$

The term $([Q_{jj} - Q_{ji}])$ can be similarly decomposed as:

$$Q_{jj} - Q_{ji} = \sum_{k=0}^{n-1}[\varepsilon(\mathcal{F}_k^{jj}) - \varepsilon(\mathcal{F}_k^{ji})] / \sum_{k=0}^{n-1} \varepsilon(\mathcal{F}_k)$$

$$= \frac{[\varepsilon(\mathcal{F}_0^{jj}) - \varepsilon(\mathcal{F}_0^{ji})] + [\varepsilon(\mathcal{F}_1^{jj}) - \varepsilon(\mathcal{F}_1^{ji})] + ... + [\varepsilon(\mathcal{F}_{n-2}^{jj}) - \varepsilon(\mathcal{F}_{n-2}^{ji})] + [\varepsilon(\mathcal{F}_{n-1}^{jj}) - \varepsilon(\mathcal{F}_{n-1}^{ji})]}{\sum_{k=0}^{n-1} \varepsilon(\mathcal{F}_k)}$$

The denominator $\sum_{k=0}^{n-1} \varepsilon(\mathcal{F}_k)$ is an invariant for a given graph; and plays no role in determining the relativities across the members of the set $E(G)$. The numerator, on the other hand, certainly does. We go over each term in the numerator, backwards, starting from $[\varepsilon(\mathcal{F}_{n-1}^{ii}) - \varepsilon(\mathcal{F}_{n-1}^{ij})]$ and going all the way to $[\varepsilon(\mathcal{F}_0^{ii}) - \varepsilon(\mathcal{F}_0^{ij})]$ which is a constant ($= 1$).

#### 3.3.1 Contribution of spanning rooted forests with $(n-1)$ edges to $\Delta_{ij}$

Consider an individual spanning rooted forest $F_{n-1}^{ii} \in \mathcal{F}_{n-1}^{ii}$: a spanning forest with $v_i \in V(G)$ as the root of the tree in which it belongs, and $(n-1)$ edges. Clearly, $F_{n-1}^{ii}$ is a spanning tree of $G$; since, $F_{n-1}^{ii}$ is spanning, acyclic and has (n-1) edges - with the only difference being that $v_i$ is distinguished as the root of this tree (say, colored *red*). Therefore:

$$\varepsilon(\mathcal{F}_{n-1}^{ii}) = |\mathcal{T}(G)| \quad (12)$$

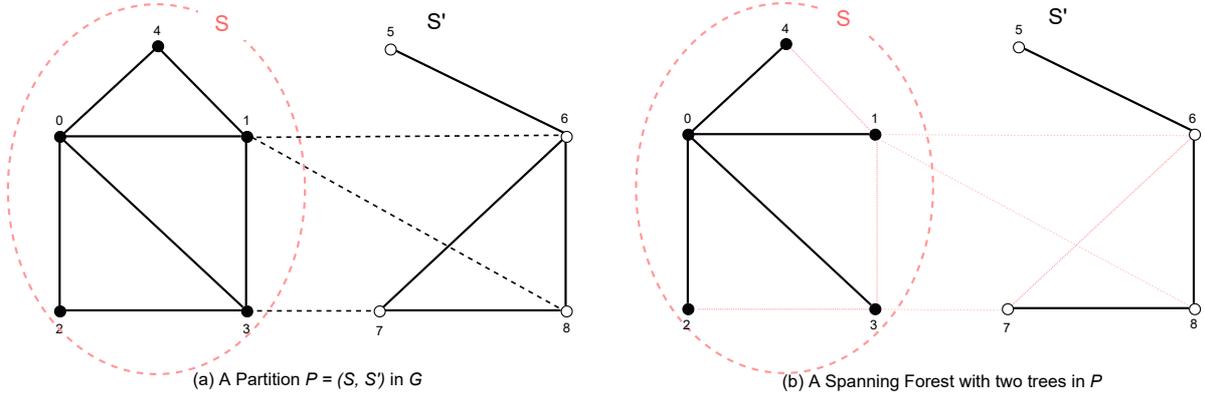

Figure 3: **Bi-Partition of a graph** $G(V, E, W)$: (a) $G(V, E, W)$ has nine vertices ($v_0$ to $v_8$) and fourteen edges $E(G) = \{e_{0,1}, e_{0,2}, e_{0,3}, e_{0,4}, e_{1,3}, e_{1,4}, e_{1,6}, e_{1,8}, e_{2,3}, e_{3,7}, e_{5,6}, e_{6,7}, e_{6,8}, e_{7,8}\}$. $S$ and $S'$ are the two sub-graphs of $G$ that constitute the bi-partition $P = (S, S')$, shown in (a), constructed by deleting edges $\{e_{1,6}, e_{1,8}, e_{3,7}\}$. $V(S) = \{v_0, v_1, v_2, v_3, v_4\}$ and $V(S') = \{v_5, v_6, v_7, v_8\}$; $E(S) = \{e_{0,1}, e_{0,2}, e_{0,3}, e_{0,4}, e_{1,3}, e_{1,4}, e_{2,3}\}$, $E(S') = \{e_{5,6}, e_{6,7}, e_{6,8}, e_{7,8}\}$ and $E(S, S') = \{e_{1,6}, e_{1,8}, e_{3,7}\}$. (b) $[T(S), T(S')]$ represent one spanning forest of $G$, subtended by the partition $P = (S, S')$. The edges in $T(S)$ are $\{e_{0,1}, e_{0,2}, e_{0,3}, e_{0,4}\}$ while those in $T(S')$ are $\{e_{5,6}, e_{6,8}, e_{7,8}\}$. The number of spanning rooted forests represented in this configuration $[T(S), T(S')]$ in (b), for the members of the sub-graph $S$ is 4, while those for the members of the sub-graph $S'$ is 5 (c.f. [41]).

where $\mathcal{T}(G)$ is the set of all the spanning trees in $G$ and $|\mathcal{T}(G)|$ is their count. Note also, that in a spanning tree, all the vertices of $G$ are present by definition. So,

$$\varepsilon(\mathcal{F}_{n-1}^{ij}) = |\mathcal{T}(G)| \tag{13}$$

Therefore, $[\varepsilon(\mathcal{F}_{n-1}^{ii}) - \varepsilon(\mathcal{F}_{n-1}^{ij})] = 0$. By symmetry, $[\varepsilon(\mathcal{F}_{n-1}^{jj}) - \varepsilon(\mathcal{F}_{n-1}^{ji})] = 0$. So, the set $\mathcal{F}_{n-1}$ does not contribute to the numerator of $\Delta_{ij}$ at all.

### 3.3.2 Contribution of spanning rooted forests with $(n-2)$ edges to $\Delta_{ij}$

In order to analyze the counts $[\varepsilon(\mathcal{F}_{n-2}^{ii}), \varepsilon(\mathcal{F}_{n-2}^{ij}), \varepsilon(\mathcal{F}_{n-2}^{jj}) \& \varepsilon(\mathcal{F}_{n-2}^{ji})]$, we need to revisit the *bi-partitions* of a graph introduced in [41].

**Definition 6** *Bi-partition $(P = (S, S'))$: A cut of the graph $G$ which contains exactly two mutually exclusive and exhaustive connected sub-graphs $S$ and $S'$ (c.f. Figure 3).*

Let, $V(S)$ and $V(S')$ be the mutually exclusive and exhaustive subsets of $V(G)$, $E(S)$ and $E(S')$, the sets of edges in the respective components $S$ and $S'$ of $P$ and $E(S, S')$, the set of edges that *violate* $P$ i.e. have one end in $S$ and the other in $S'$. Also, let $\mathcal{T}(S)$ and $\mathcal{T}(S')$ be the set of spanning trees in the respective component sets $S$ and $S'$. We denote by $\mathcal{P}(G)$, the set of all bi-partitions of $G(V, E)$. Clearly, a given $P = (S, S')$ represents a sub-graph from which $E(S, S')$ have edges have been deleted. A vertex $v_i \in V(S)$ stays connected to $|V(S)| - 1$ other vertices and gets disconnected from $|V(S')|$ vertices; while a vertex $v_j \in V(S')$ stays connected to $|V(S')| - 1$ other vertices and gets disconnected from $|V(S)|$ vertices. For convenience, we will always call the sub-graph in which $v_i$ is as $S$. Then:

$$\varepsilon(\mathcal{F}_{n-2}^{ii})_{|P(S,S')} = |\mathcal{T}(S)||\mathcal{T}(S')||V(S')| \qquad v_i \in V(S) \tag{14}$$

and,

$$\begin{aligned}\varepsilon(\mathcal{F}_{n-2}^{ij})_{|P(S,S')} &= |\mathcal{T}(S)||\mathcal{T}(S')||V(S')| & if\ v_j \in V(S),\\ &= 0 & otherwise.\end{aligned}$$

So:

$$\varepsilon(\mathcal{F}_{n-2}^{ii}) - \varepsilon(\mathcal{F}_{n-2}^{ij}) = \sum_{P \in \mathcal{P}(G)}^{v_i \in V(S), v_j \in V(S')} |\mathcal{T}(S)||\mathcal{T}(S')||V(S')| \tag{15}$$

Note that the condition $v_i \in V(S)$ & $v_j \in V(S')$ maps to those partitions in $\mathcal{P}(G)$ in which the edge $e_{ij} \in E(S, S')$.

By symmetry:

$$\varepsilon(\mathcal{F}_{n-2}^{jj}) - \varepsilon(\mathcal{F}_{n-2}^{ji}) = \sum_{P \in \mathcal{P}(G)}^{v_j \in V(S'), v_i \in V(S)} |\mathcal{T}(S)||\mathcal{T}(S')||V(S)| \quad (16)$$

which yields:

$$\varepsilon(\mathcal{F}_{n-2}^{ii}) - \varepsilon(\mathcal{F}_{n-2}^{ij}) + \varepsilon(\mathcal{F}_{n-2}^{jj}) - \varepsilon(\mathcal{F}_{n-2}^{ji}) = \sum_{P \in \mathcal{P}(G)}^{v_i \in V(S), v_j \in V(S')} |\mathcal{T}(S)||\mathcal{T}(S')|(|V(S')| + |V(S)|)$$

$$= n \sum_{P \in \mathcal{P}(G)}^{v_i \in V(S), v_j \in V(S')} |\mathcal{T}(S)||\mathcal{T}(S')|$$

The final term in the $RHS$ above is of great interest. Henceforth, we denote it by $\tau_{ij}$ for convenience. Given that $e_{ij}$ is a cut edge in $E(S, S')$, it is easy to see that:

a. **Handling Bridge Edges**: If $e_{ij}$ is a *bridge* edge in a graph $G$, then there is only one bi-partition $P$ in which it is present in the set $E(S, S')$. Algorithm MCST/LCST never deletes it, anyway.

b. **Contribution to Connectivity between vertices in** $S$ & $S'$: No edge in $E(S, S')$ contributes to inter-vertex connectivity or to the spanning trees defined over the vertices of the sub-graph $S$, or the vertices of the sub-graph $S'$. $|\mathcal{T}(S)|$ is a measure of connectivity in the sub-graph $S$; and $|\mathcal{T}(S')|$ a measure of connectivity in the sub-graph $S'$. Higher the value of $|\mathcal{T}(S)|$ and $|\mathcal{T}(S')|$, shorter the inter-vertex distances in $S$ and $S'$ respectively (owing to greater path diversity); and lower the impact of the deletion of $e_{ij}$ in inter-vertex path diversity. By extension, higher the value of $\tau_{ij}$, lower the impact of a potential deletion of $e_{ij}$ from the graph during an iteration of the algorithm MCST.

c. **Contribution to Connectivity in Spanning Trees**: If we choose a $T(S) \in \mathcal{T}(S)$, an edge in $E(S, S')$ and a $T(S') \in \mathcal{T}(S')$, together they form a spanning tree $T(G) \in \mathcal{T}(G)$. In fact, all the spanning trees of $G$ can be obtained by combining elements from $\mathcal{T}(S), \mathcal{T}(S')$ & $E(S, S')$ for any given partition $P \in \mathcal{P}(G)$. Note, as we only regress over $e_{ij}$ that are not bridge edges, each $e_{ij} \in E(S, S')$ is *covered* by at least one other edge in $E(S, S')$ which can be used as an alternative to construct spanning trees across the bi-partition $P$. Higher the value of $\tau_{ij}$, greater the fraction of spanning trees of $G$ in which $e_{ij}$ is covered by at least one other edge; and its deletion has a lower impact.

Analogously, for LCST, edges with lower $\tau_{ij}$ are suitable candidates. The rest of the argument follows *as is*.

### 3.3.3 Contribution of spanning rooted forests with $(n-3)$ edges to $\Delta_{ij}$

Extending the argument to *tri-partitions* (c.f. Fig. 4): three mutually exclusive and exhaustive sub-graphs of $G$, denoted by $S, S'$ and $S''$, we obtain the following:

$$\varepsilon(\mathcal{F}_{n-3}^{ii}) - \varepsilon(\mathcal{F}_{n-3}^{ij}) + \varepsilon(\mathcal{F}_{n-3}^{jj}) - \varepsilon(\mathcal{F}_{n-3}^{ji}) = n \sum_{P \in \mathcal{P}(G)}^{v_i \in V(S), v_j \in V(S')} |\mathcal{T}(S)||\mathcal{T}(S')||\mathcal{T}(S'')| \quad (17)$$

By abuse of notation, $P = (S, S', S'')$ is a tri-partition in this case. All the definitions from the previous section follow as is: with $V(S), V(S'), V(S'')$ representing the sets of vertices in each sub-graph, $E(S), E(S'), E(S'')$ represent the edges in the respective sub-graphs, and $E(S, S'), E(S, S''), E(S', S'')$ represent the edges between the sub-graphs. In fact, the argument is easily generalized for any value of $(n - \kappa)$ where $\kappa = \{2, 3, 4, ..., n\}$.

### 3.3.4 Contribution of spanning rooted forests with $1$ edge to $\Delta_{ij}$

Clearly, $\varepsilon(\mathcal{F}_1^{ii}) = d(v_i) + 2 \cdot (|E(G)| - d(v_i)) = 2 \cdot m - d(v_i)$, where $d(v_i)$ is the degree of the vertex $v_i$, and $m = |E(G)|$ is the total number of edges in $G$. Similarly, $\varepsilon(\mathcal{F}_1^{ij}) = 1$ if $e_{ij} \in E(G)$, 0 otherwise. As we are only concerned with vertex pairs $(i, j) \in V(G) \times V(G) : e_{ij} \in E(G)$, we obtain:

$$[\varepsilon(\mathcal{F}_1^{ii}) - \varepsilon(\mathcal{F}_1^{ij})] + [\varepsilon(\mathcal{F}_1^{jj}) - \varepsilon(\mathcal{F}_1^{ji})] = [2 \cdot m - d(v_i) - 1] + [2 \cdot m - d(v_j) - 1]$$
$$= 4 \cdot m - [d(v_i) + d(v_j)] - 2$$

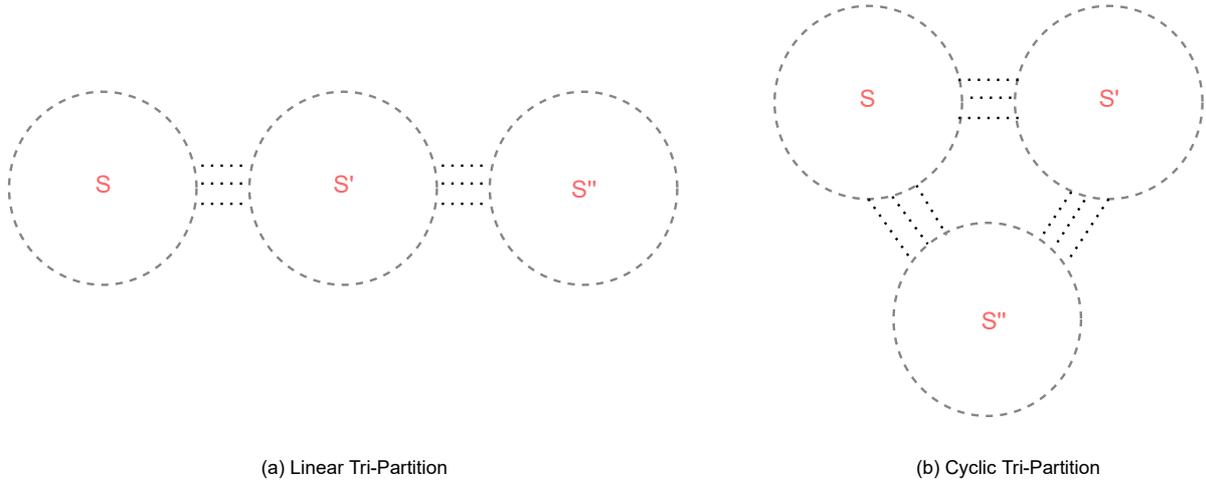

(a) Linear Tri-Partition

(b) Cyclic Tri-Partition

Figure 4: **Tri-Partitions of a graph** $G(V, E, W)$: A schematic showing the linear and cyclic variants. The labels $\{S, S', S''\}$ are arbitrarily assigned here. From the point of view of an edge, it can only belong to one of the sets: $E(S), E(S'), E(S''), E(S, S'), E(S', S'')$ in the linear variant; and $E(S), E(S'), E(S''), E(S, S'), E(S', S''), E(S'', S)$ in the cyclic variant. It is easy to see that the linear configuration in (a) is simply a special case of the more general cyclic configuration in (b); induced by the condition $E(S, S'') = \phi$. Higher order partitions can similarly have multiple variants depending upon the structure of the graph.

### 3.3.5 Contribution of spanning rooted forests with 0 edges to $\Delta_{ij}$

Finally, we evaluate the term $[\varepsilon(\mathcal{F}_0^{ii}) - \varepsilon(\mathcal{F}_0^{ij})]$ and $[\varepsilon(\mathcal{F}_0^{jj}) - \varepsilon(\mathcal{F}_0^{ji})]$. There can only be one spanning rooted forest of $G$ with 0 edges: a forest in which all the $n$ vertices of $G$ are isolated, single-vertex trivial trees and serve as roots. Clearly, in a forest of this kind, the question of $(i, j)$ being in the same tree, doesn't arise. Hence, $\forall i \in V(G) : \varepsilon(\mathcal{F}_0^{ii}) = 1$ and $\forall (i, j) \in V(G) \times V(G) : \varepsilon(\mathcal{F}_0^{ij}) = 0$. Hence, the term contributes exactly $+2$ to $\Delta_{ij}$ for any pair $(i, j)$: $+1$ for $v_i$ and $+1$ for $v_j$; and is an invariant.

To summarize, therefore, the edge $e_{ij}$ with the maximum $\Delta_{ij}$ in a graph contributes the least inter-vertex connectivity and represents an ideal candidate for elimination during a regress step during the MCST algorithm in Table 1. This is owing to the fact that the value itself is based on the number of spanning trees that remain in tact in the graph despite the removal of the $e_{ij}$. It is easy to see that the same argument follows in the case of LCST construction.

## 4 Empirical Evaluation

### 4.1 The Complete Graph $K_n$ of Order $n$

The complete graph $K_n$, by definition contains every graph of order $n$ as a sub-graph in it. Hence, all possible trees of order $n$ are sub-graphs of $K_n$ as well. If the algorithm in Table 1 indeed generates an MCST ($T^*(G)$) - or, an LCST ($T^\#(G)$), based on the choice of the extremality criteria - we would expect it to return an $S_n$ - or, a $P_n$ - in step 6 for any $G(V, E, W) = K_n$ that is input in step 1. Figure 5 demonstrates that this is indeed the case for $K_4$. When the MCST and LCST variants are run over $K_4$, we attain the expected output. Following are worth noting:

a. **Number of Iterations**: Both sequences $\{(a) \to (d)\}$ and $\{(e) \to (h)\}$ conclude in three regress steps each; leaving behind exactly three *bridge* edges connecting the four vertices [4].

b. **Maintaining Connectedness**: The intermediary graphs always stay connected. The bridge edge $e_{0,3}$ in $(c)$ is retained in the regress $(c) \to (d)$ despite of having the highest $\Delta_{0,3} = 0.6$, amongst all the edges in $(c)$. The edge $e_{1,2}$ is deleted instead, which has the highest $\Delta_{1,2} = 0.5$, amongst the *non-bridge* edges.

c. **Breaking Ties**: When more than one edge has the maximum (MCST) - or, minimum (LCST) - value of $\Delta_{ij}$ in an iteration, any one - and exactly one - of them is deleted.

---

[4]Equivalently, the algorithm can be concluded when $\forall (i, j) \in V(G) \times V(G) \, \& \, e_{ij} \in E(G) : \Omega_{ij} = 1$.

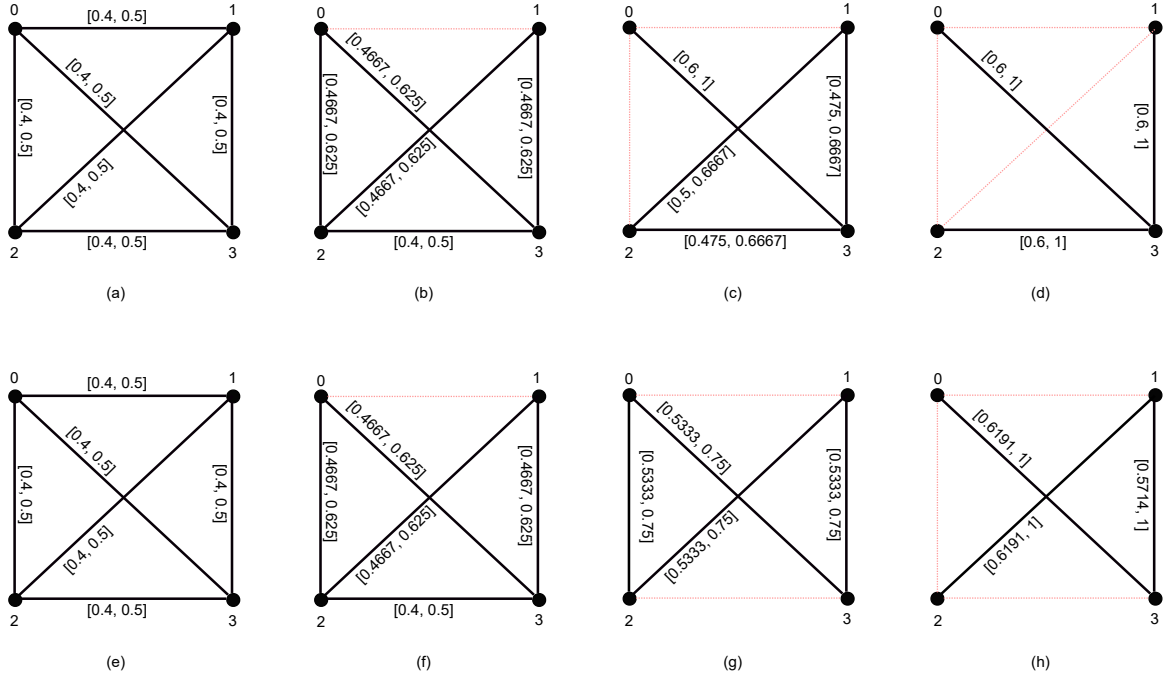

Figure 5: **Algorithm MCST and LCST over** $K_4$: Sequence $(a) \to (d)$ represents the iteratively greedy MCST extraction; while, sequence $(e) \to (h)$ represents the iteratively greedy LCST extraction. Note that the sub-graph in (d) and (h) are indeed $S_4$ (one of the four possible) and $P_4$ (one of the twelve possible). The tuple on the edges represents the pair $[\Delta_{ij}, \Omega_{ij}]$ for the edge $e_{ij}$ in each iteration. Red dotted edges represent all the deleted edges until a given iteration.

But $K_4$ is a special graph with only $S_4$ (four different trees) and $P_4$ (twelve different trees) as its spanning trees. In order to ensure that this is indeed the case, we repeat the experiment for $K_n : n = 5, 6, 7, 8, 9, 10$; and each time obtain $S_n$ and $P_n$ as the end result. Note, we are aware that these illustrations over $K_n$ are necessary to support our claims, but not sufficient. They are simply meant to be basic sanity checks.

### 4.2 The Erdős-Renyi Random Graphs

Next, we generate over 500 Erdős-Renyi random graphs (ER-graphs) [14] of varying orders: $n \in [10, 99]$ and edge densities $\rho \in [0.25, 0.50]$. ER-graphs are considered as benchmarks in many studies in literature and hence make for a natural choice in ours too. For each graph in this set, we generate:

a. One MCST ($T^*(G)$) and LCST ($T^\#(G)$) pair.

b. A set of random spanning trees [9, 16, 49]. We use the version in [49] to generate $n$ random spanning trees for a graph of order $n$ (by starting the algorithm once from each vertex).

Figure 6 shows comparative results on the average inter-vertex shortest path distances in the original graph ($G$), the MCST ($T^*(G)$), the LCST ($T^\#(G)$) and the mean across all random spanning trees generated (c.f. [b.] above). The graphs are arranged first by their respective orders in ascending order and then for each order, by ascending orders of the edge density ($\rho$). The results validate our claim that the MCST and LCST produced by our algorithm indeed are *extremal* in compactness. As one would expect, minor variations in absolute values do occur when $n$ is kept fixed and $\rho$ is varied; but these do not seem to change the overall pattern; the relativity of compactness between the various tree classes (c.f. APPENDIX for a discussion on a couple of minor exceptions). For completeness, we also study the diameters of the trees thus produced (c.f. Figure 6(b)), and notice the same general trends.

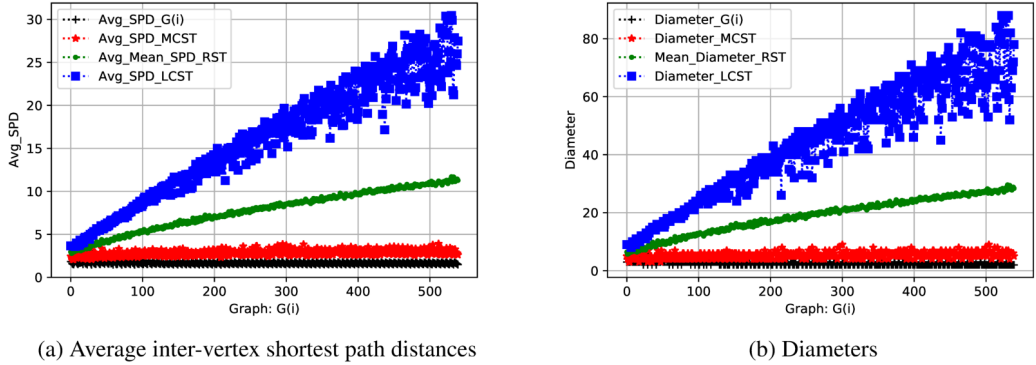

(a) Average inter-vertex shortest path distances  (b) Diameters

Figure 6: **Algorithm MCST and LCST over a set of ER-graphs**: (a) average shortest path distances in $G$, MCST, LCST and mean of average shortest path distance across all the random spanning trees, (b) diameter of $G$, MCST, LCST and mean diameter across all the random spanning trees. The number of random spanning trees for each graph $G$ is $n = |V(G)|$; we generate one random spanning tree per vertex in $V(G)$ for this analysis.

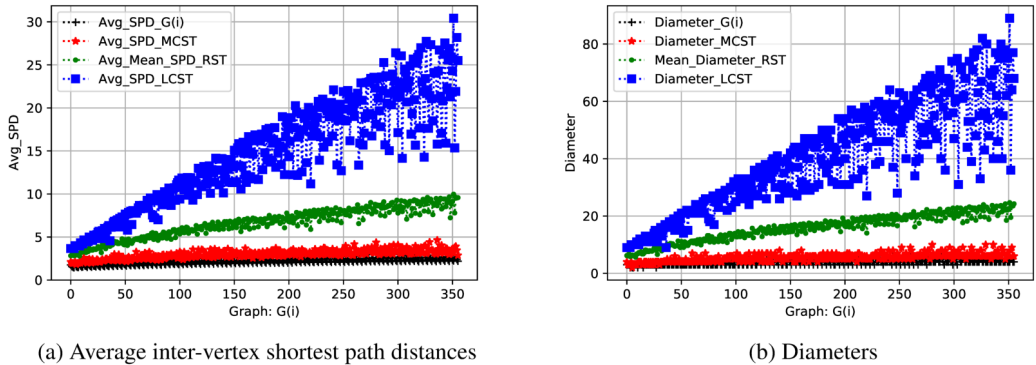

(a) Average inter-vertex shortest path distances  (b) Diameters

Figure 7: **Algorithm MCST and LCST over a set of BA-graphs**: (a) average shortest path distances in $G$, MCST, LCST and mean of average shortest path distance across all the random spanning trees, (b) diameter of $G$, MCST, LCST and mean diameter across all the random spanning trees. The number of random spanning trees for each graph $G$ is $n = |V(G)|$; we generate one random spanning tree per vertex in $V(G)$ for this analysis.

### 4.3 The Barabási-Albert Scale-Free Graphs

Finally, we look at the performance of the MCST and LCST algorithms over a set of about 360 Barabási-Albert scale-free graphs (BA-graph) [8, 10, 24] of varying orders and edge densities: $n \in [10, 99]$ and the average degree per node varying between $[2, 5]$. A BA-graph is a result of the so called *preferential attachment* generative process in which each in-coming node attaches itself with pre-existing nodes in the erstwhile graph with a constant number of edges, and the *preference* of attachment is proportional to the node degrees of the other vertices. This class of graphs also finds widespread use in literature; and hence makes for an important baseline class to validate our work on. Once again, the compactness relativities hold between the graph, the MCST, the LCST and the random spanning trees; both for the average inter-vertex shortest path distances as well as the diameters. We note that as average vertex degree increases (for the same order), the gap between the three classes - the MCST, the LCST and the random spanning trees - does seem to reduce. One would expect this given the so called *small-world* nature of the BA-graphs (c.f. Figure 7). This is largely due to short diameters which in turn is owed to the *power-law scale-free* degree distributions and the existence of *rich club connectivity* (RCC) by the virtue of which a few *influential* vertices (vertices of high degrees), are inter-connected with each other with high probability, and also provide connectivity between any arbitrary vertex-pair. We refer interested readers to the original sources [8, 10, 24] and the references therein, for details.

## 5 Complexity Analysis

Complexity analysis of the algorithm in Table 1 is relatively straightforward. The **while** loop in step 2 runs $m - (n - 1)$ times; $m$ being the total number of edges and $n$, the order of the original $G$ (number of nodes). So, the computational complexity of the outer loop is $O(m)$. Each **while** iteration, in turn, involves the computation of a matrix inverse ($Q$) and a pseudo-inverse ($L^+$). These matrices belong to $\Re^{n \times n}$; so, both operations have $O(n^3)$ complexity [29]. In addition, the **for each** loop within the **while** runs over the set of edges again; and has an $O(m)$ complexity again. Overall complexity of the algorithm is:

$$O(m * [(2 \cdot n^3) + m]) = O((2 \cdot m \cdot n^3) + (m^2)) = O(m \cdot n^3) \qquad (18)$$

Therefore, our proposed *iteratively greedy rank-and-regress* algorithm is still polynomial time; albeit a relatively higher order polynomial. In the worst case, $m = O(n^2)$, which yields a further reduction of the form in (18) to $O(n^5)$. For most real world graphs, $m$ is often $O(n)$ as opposed to $O(n^2)$; which reduces the complexity by an order of magnitude to $O(n^4)$ for real world graphs.

Further reduction in time complexity can be attained by using the incremental computation methods for $Q$ and $L^+$ [18, 22, 42]; as well as, hardware accelerators (e.g. GPUs) [44, 47].

## 6 Related Work

The literature on spanning trees of graphs is vast. While it is impossible - or, at least, impractical - to provide a comprehensive survey, we give a brief summary of some of the most commonly used variants here. As stated earlier in this work, *Minimum Spanning Trees* (MST) are the most pervasively known class of spanning trees; and the *Prim's* and *Kruskal's* algorithms the most widely used for constructing MSTs [48]. The MST is, alas, most meaningful for weighted graphs; and, at best, of limited use in the unweighted case. Another class of spanning trees are the shortest path trees [1] which try to construct compact spanning trees starting from a particular vertex. The spanning trees thus generated are not compact from the standpoint of the graph; and have found much use in fields of communication networks and distributed systems [2, 4]. For unweighted graphs, *breadth first* and *depth first* searches can also be used to generate spanning trees of extremal compactness from the standpoint of a vertex (the root) [19, 48]. Once again, these are not designed for - nor do they attain - desired compactness from the point of view of the graph itself. A population of random spanning trees of the graph is indeed more suitable for a statistical study. The most prominent work on random spanning trees is due to Aldous and Border, who individually studied this problem [9, 16].

Similar to the study of spanning trees, there is a thriving tradition in literature of extracting other sub-graphs of desired compactness from a graph. We refer the reader to [15], and the references there in, where the authors propound on graph *sparsification* and *coarsening* techniques and present a unified theory of *sparsification* and *coarsening*. Antecedents of such work are also found in [25, 26, 27, 39, 40].

And the world of directed graphs is not entirely untouched from this sort of activity either. All the aforementioned algorithms have some or the other analogue for the case of directed graphs. The spanning tree, in the case of a digraph, simply becomes a spanning rooted *arborescence*. Much of the algebraic and topological/graph-theoretic paraphernalia extends to the world of digraphs [6, 7, 13].

## 7 Conclusion and Future Work

In this work, we introduced the concept of *Most, and Least, Compact Spanning Trees - viz.* $T^*(G)$ and $T^\#(G)$ - of a simple, connected, undirected and unweighted graph $G(V, E, W)$. We then presented an *iteratively greedy rank-and-regress* method that produces at least one $T^*(G)$ or one $T^\#(G)$ by eliminating one extremal edge per iteration. The rank function for performing the elimination is based on making a greedy choice over the elements of the matrix of *relative forest accessibilities* ($Q$) of a graph and the related *forest distance* associated with the edges of $G$. The algorithm guarantees polynomial time convergence ($O(m \cdot n^3)$) in the general case, and ensures that at all intermediary stages of the *regress*, the graph stays connected. We provided empirical evidence in support of our methodology; and discussed the computational efficiencies that enable extraction of $T^*(G)$ and $T^\#(G)$ within polynomial times.

While useful for practical purposes already, establishing formal bounds on the level of compactness produced by our proposed algorithm - akin to those available for random spanning trees [23] - is desirable. Also, the question of extending the proposed methodology to the case of directed graphs - or, digraphs - is tantalizing in its own right. Indeed, there exists an analogous matrix of *relative forest accessibilities* for the directed case [6, 7, 13]. We motivate all these for future work.

# Acknowledgment

The authors would like to thank Dr. Daniel Boley and Dr. Zhi-Li Zhang for preliminary discussions related to this work.

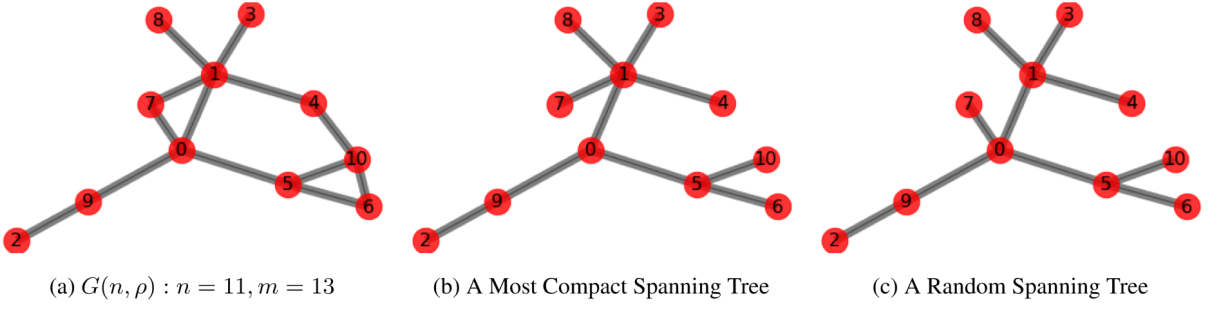

(a) $G(n, \rho) : n = 11, m = 13$  (b) A Most Compact Spanning Tree  (c) A Random Spanning Tree

Figure 8: **ER-Graph I**: The MSCT and RST differ in exactly one edge ($e_{1,7}$ in the MCST vs. $e_{0,7}$ in the RST). The resulting discrepancy in the average shortest path lengths is negligible for all practical purposes ($\approx 1.2\%$). The diameters of all three - $G$, MCST and RST - are exactly the same ($= 4$).

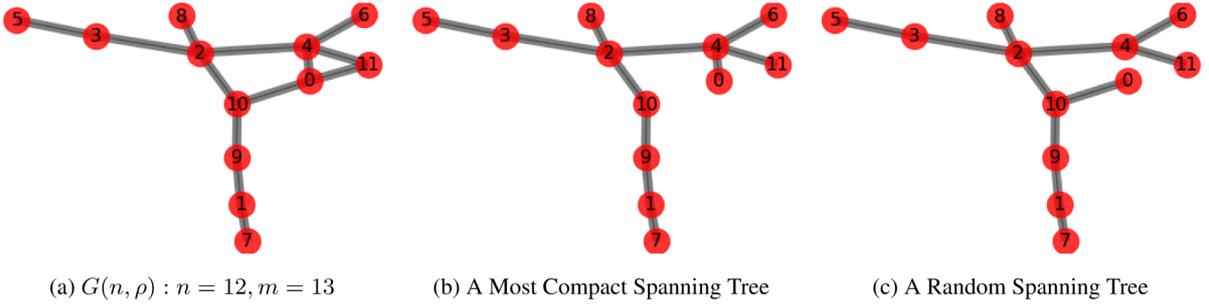

(a) $G(n, \rho) : n = 12, m = 13$  (b) A Most Compact Spanning Tree  (c) A Random Spanning Tree

Figure 9: **ER-Graph II**: Once again, the MCST and the RST differ in exactly one edge ($e_{0,4}$ in the MCST vs. $e_{0,10}$ in the RST). The discrepancy in this case is even smaller ($\approx 1\%$). The diameters in this case are also the same for all three ($= 6$).

## APPENDIX

**A Couple of Exceptions**

During the experiments on ER-Graphs (c.f. §4.2), we encountered two samples where the MCST had a higher value for the average inter-vertex shortest path lengths than at least one of the random spanning trees. Figures 8 and 9, show the respective graphs and their spanning trees. The difference between the average shortest path lengths (compactness) in the MCST and one of the random trees in both cases is negligible. Nevertheless, it is there to be seen; and hence must be accounted for. Hence, we motivate obtaining formal bounds on the output of our algorithm - akin to those for random spanning trees [23] - in future work.